# Managing FAIR Knowledge Graphs as Polyglot Data End Points: A Benchmark based on the rdf2pg Framework and Plant Biology Data

Marco Brandizi [a,*,†], Carlos Bobed [b,c,*], Luca Garulli [d], Arné de Klerk [a], Keywan Hassani-Pak [a]
[a] *Rothamsted Research, Harpenden, AL5 2JQ, UK*
[b] *Aragon Institute of Engineering Research (I3A)*
[c] *University of Zaragoza, Zaragoza, Spain*
[d] *Arcade Data Ltd, London, UK*
[*] *These authors contributed equally*
[†] *Corresponding author, email: marco.brandizi@rothamsted.ac.uk*

**Abstract.** Linked data and labelled property graphs (LPG) are two data management approaches with complementary strengths and weaknesses, making their integration beneficial for sharing datasets and supporting software ecosystems. In this paper, we introduce rdf2pg, an extensible framework for mapping RDF data to semantically equivalent LPG formats and databases. Utilising this framework, we perform a comparative analysis of three popular graph databases - Virtuoso, Neo4j, and ArcadeDB - and the well-known graph query languages SPARQL, Cypher, and Gremlin. Our qualitative and quantitative assessments underline the strengths and limitations of these graph database technologies. Additionally, we highlight the potential of rdf2pg as a versatile tool for enabling polyglot access to knowledge graphs, aligning with established standards of linked data and the semantic web.

## 1. Introduction

While the Semantic Web and ontology engineering are still fundamental as common languages to exchange data (mainly related to the 'Interoperable' of the FAIR principles [1]), we have seen some different recent trends in networked and shared knowledge, reflecting the difficulties that practitioners still experience with dealing with these approaches. For instance, while domains like life science still benefit from precise annotations based on OWL ontologies [2], complementary models such as schema.org and Bioschemas [3] are becoming popular as 'lightweight' data models, which are easier to use and fit well use cases such as harmonising large and heterogeneous datasets or making them visible on search engines and accessible via the Web [4].

Another emerging trend, which has arisen in domains like data intelligence or machine learning, is the adoption of labelled property graphs (LPG), which are the basis of graph databases [5] and other frameworks [6,7]. With respect to RDF triples, the LPG graphs are less fine-grained, being able to keep together an entity's properties in a single node, and most importantly, they are more expressive in representing 'binary relations with annotations' (see Figure 1). The proliferation of LPG-based technologies has stimulated collective efforts to standardise (again…) at least the query languages, with examples like Open Cypher [8], later turned into GQL [9], or the Gremlin/TinkerPop framework [10].

In the authors' experience, LPG-based data management is not an alternative to more 'traditional' Linked Data approaches, as the two have a set of complementary advantages and disadvantages [11]. For instance, while systems such as Neo4j [12] can offer fast and scalable access to LPG knowledge graphs by means of the expressive Cypher language, they do not have reference/standard formalisms for data exchange, nor does a standard exist to represent LPG schemas and advanced schema-related entities, such as ontologies. Both these aspects are the focus of the Semantic Web stack and existing literature shows that these two approaches can be usefully integrated [11,13,14].

A consequence of this is the need for bridging these two worlds, in order to obtain good integrations and benefit from the best of the two. In this paper, we present rdf2pg, an extensible framework to convert RDF graph data into various target LPG graph formats and graph databases by means of a user-defined mapping between the source RDF data and the target LPG. As we will show, this has the main advantage of allowing for a sensible and domain-aware mapping between these two graph data models that, though very similar, have significant 'natural' ways to represent certain semantics (e.g., relations with properties vs reified statements). Another objective of this paper is to explore, both qualitatively and quantitatively, the use of different graph databases and query languages, when used to store data that are conceptually the same and are aligned to the same conceptual model. We show how our rdf2pg tool makes such alignment possible and we base our analysis on plant biology datasets, which is a typical use case for which the knowledge graph model fits very well [15–17].

*1.1. Motivational use cases*

In this section, we describe the use cases that motivated the development of the rdf2pg tool and the management of respective datasets with graph databases and their query languages.

**KnetMiner**. The rdf2pg framework is an evolution of the rdf2neo tool [13]. The need for the latter was born within the KnetMiner project [17,18], a platform that provides functionality to explore knowledge about molecular biology. KnetMiner offers the end users an easy-to-use web application to quickly search genes of interest and related entities (e.g., diseases, biological processes, scientific literature), as well as to visualise them in various forms, including knowledge networks. This is based on knowledge graphs that are built starting by integrating many useful public data sources. The same data that powers the web application are also available programmatically, in the form of a specific web API [19], a SPARQL endpoint, and a Neo4j data endpoint (that is, Cypher access via the BOLT protocol and the Neo4j browser). The KnetMiner team decided to use (and share) both RDF-based data and Neo4j due to the complementary pros and cons that the two have, both in general and for the specific platform needs. For example, after many years of varying success, RDF and the Semantic Web stack are still reliable standards to integrate data, being particularly suitable for sharing schemas and ontologies [20] and for operations such as automatically merging datasets referring to the same Uniform Resource Identifiers (URIs), ontology terms and public identifiers. On the other hand, many developers and bioinformaticians find Cypher and Neo4j an easier technology to work with. As an example of the latter, they use Cypher to define 'semantic motifs', which are graph patterns capturing chains from genes to relevant entities [21] (e.g., gene->protein->bio-process->article-mention), so that the related entities can be exploited to realise application-level functionality like displaying semantic associations between genes of interest. Another advantage given by Neo4j is that its ecosystem offers useful tools to manage and analyse the data, such as Bloom [22] or NeoDash [23]. Recently, the latter has been used to summarise the general characteristics of KnetMiner datasets [24,25].

**Enterprise Semantic ETL**. This use case deals with the integration of data coming from different information systems within different enterprise domains, including health and insurance, realised by NTT Data in collaboration with the University of Zaragoza, within the project 'Semantic Data Management in Knowler'. Here, we describe the main project aspects by considering the non-disclosure requirements set by the companies involved. One of the seed domains was the integration of all the in-house information about the ongoing projects, the employees and their skills. Such information was scattered across various underlying information systems (usually following different governance policies). To integrate such information, a knowledge graph was built out from many different structured sources, designing a Semantic ETL (Extract-

Transform-Load) pipeline. Each source (usually structured) followed a different model and, in order to include everything together under RDF format, many times reification of non-binary relationships must be applied. In our initial domain, we had *Employees* related to their *Skills* (both soft and hard skills, such as *Technologies*) and their particular levels achieved (evaluated within the company). Thus, in RDF, we needed to reify such relationships in order not to lose any information in the integration process. Moreover, once we have the triple store completely populated, it was especially convenient to have all the flexibility required to build views on property graphs, possibly materialising traversals over different property paths. For example, we could materialise the time that an employee has been working on a particular kind of project and directly include that in the property graph appropriately. In this ETL, we used an ontology as an integration umbrella to bring together data from different sources. For information integration purposes, we used RDF triple stores, which are particularly effective at handling Linked Data [26,27]. For other tasks, such as building an adaptive presentation layer, we used property graph databases, mainly because internal tests showed their better performance. We designed the pipeline in a way that ensures high flexibility in building the presentation views. Moreover, we also paid attention to the particular requirements in enterprise scenarios, where software licence costs and avoidance of vendor-locking are two factors which might lead to project failure. Those two risks were avoided by the adoption of standard formats and by using the property graph as a general and unified data model. We chose GraphML as the reference format to produce LPG data in the ETL pipeline as it allowed us to store labelled multigraphs in an extensible way (adding attributes or references to the edges themselves) and it is supported by many different graph databases and visualisation tools.

In both KnetMiner and Semantic ETLs, an approach and software tool are needed to align RDF-represented data with different LPG data. Initially, the rdf2neo framework was developed [13], which offered the feature to map RDF data models to the LPG model that is supported by the Neo4j database, and in a configurable way, which is decided by data managers and is the most appropriate for the datasets they deal with and their semantics. Later, similar needs have arisen in the Semantic ETLs scenario. This has naturally led us to extend the initial rdf2neo software into the rdf2pg infrastructure that we present hereby. As described later, we introduce the concept of mapping RDF to an abstract LPG model which allows for the execution of a specialised data generator. This same generator materialises the abstract property graph into a specific format or technology. By following this approach, we have repurposed a considerable amount of existing code for the development of an RDF->Neo4j and RDF->GraphML converter. Furthermore, the same framework remains extendable to accommodate similar or related use cases.

## 2. Technologies and methods

**The semantic web**. As well known, RDF is part of a stack of standard technologies defined under the vision of the Semantic Web [28–30]. Its main idea is to leverage the World Wide Web concepts and standards to share data in the same way we share human-targeting documents. As a result of this concept, the core of the RDF model (Figure 1, bottom) consists of binary relations between entities or about entities and data values, where the entities are described by means of resolvable URIs. URIs serve as both universal entity identifiers and, in most cases, they are web URLs that employ HTTP technology to provide documents of RDF statements about the identified entities they refer to. Over the years, a number of RDF-based formal languages (and standards) have been developed to characterise the data semantics with either lightweight schemas or advanced ontologies. As mentioned earlier, the RDF data model is very fine-grained: everything is a triple, including a property associated with an entity (such as name, surname, 'protein description'). Together with the use of URIs, this makes the merge of data about the same entities very straightforward. At the same time, in the original triple-only model it is neither possible to isolate a set of properties for an entity (eg, two pairs of name+description for the same protein, coming from two different sources), nor to associate properties (or other entities) with a triple (eg, the text mining software name and the date of a 'mentions' relation between a document and a product). Both can be obtained by modelling patterns such as RDF reification [31], which is considered a difficult to use technique. Recently, the RDF-Star extension to RDF has been proposed [32], which allows one to use triples as subjects of other triples, making RDF more similar to LPGs bringing the advantages of the latter into the Semantic Web world. We are interested in the future adoption of this approach as a standard.

Most graph database technologies have associated graph query languages. In KnetMiner, we utilise the Virtuoso triple store (a synonym for RDF-based graph database) to store our RDF data and make them publicly accessible via SPARQL, the standard query language for RDF, which is part of the Semantic Web stack [29,33]. SPARQL is essentially a graph pattern formalism for RDF, with a syntax that is a mix of the Turtle encoding format for RDF and SQL. Because the native form of the KnetMiner datasets is not based on RDF, we also employ the Jena framework, including the TDB triple store component, for software tools that convert such data into RDF [34,35]. TDB is particularly well-suited for programmatic access, while it is not the most performant SPARQL engine available, which is why we have opted for Virtuoso for the public endpoint.

**Labelled property graphs**. As mentioned above, Labelled Property Graph models are less fine-grained than RDF (Figure 1, top), meaning they support the notion of nodes and relations between nodes, both of which may have attributes attached, in the form of name/value pairs. The attribute values are usually of plain data types (e.g., string, number). Both nodes and relations usually have special additional attributes, such as, using the common jargon, 'labels' for nodes and 'types' for relations, commonly used to characterise the represented entity type. The approximate equivalent in RDF is the standard property `rdf:type` and the predicate's URI respectively, while many other models have similar concepts, e.g., class type in object-oriented languages. While in general, LPG implementations exist that support multiple relation types, here we will stick to the graph models used by Neo4j and the Gremlin framework, where only nodes can have multiple labels. As previously mentioned, RDF and LPGs are similar, though they have significant differences, especially in the details, and they have orthogonal technical advantages and disadvantages.

Currently, our rdf2pg supports the conversion of RDF to two LPG targets, the Neo4j graph database and the GraphML format. The converter for the latter has been designed and developed with a focus on populating Gremlin-compatible graph databases (ie, to load GraphML files via Gremlin commands).

Neo4j is one of the most popular graph databases built on the LPG model. It is known for its ease of deployment and maintenance, as well as its powerful query language, Cypher. Neo4j also offers a rich ecosystem of applications and tools designed to interact seamlessly with its data format, including a graph data science framework and database functions to ease graph embedding.

GraphML [36,37] is an XML-based language to define generic graphs, supporting labelled (directed, undirected, mixed) multigraphs and hypergraphs in an extensible way. Unsurprisingly, the format has elements like *<graph>*, *<node>* and *<edge>*. More-
over, it allows for the extension of the core attributes used in node and edges, by means of definitions like: *<key id = "prefName" for = "node" attr.name = "Preferred Name" attr.type = "string" />*. Attributes and *<data>* elements provide the aforementioned flexibility. Additionally, GraphML is supported by many graph tools and graph databases.

As presented on their project page, Apache TinkerPop [38] is a graph computing framework for both graph databases (OLTP) and graph analytic systems (OLAP). In essence, it is a unifying layer that offers an abstraction over the graphs and their processing, allowing different vendors to implement their specific implementations.

Gremlin [10,39,40] is the graph traversal language employed by Apache TinkerPop. Its name stems from the metaphor of a gremlin hopping from graph element to graph element while performing calculations and gathering data. This is how traversals, a central concept for data exploration and manipulation, are described in Gremlin. More formally, a traversal is a sequence of steps. For instance, Figure 5 shows a traversal that walks the composition relationship between protein complexes and the proteins they're made of. The query has basic traversal steps, plus operators similar to projection operators in other languages. In the KnetMiner use case, we have used ArcadeDB [41] to experiment with Gremlin queries on the plant biology data sets described later. ArcadeDB is a multi-model database [42], which is available as open source code, derived from the OrientDB database [43]. ArcadeDB supports, among other data models, graphs, document stores, relational/SQL databases. Although a recent and still progressing project, it is made fast by software development that exploits lower-level Java optimisations and avoids time and memory-consuming abstractions available in the standard Java language and libraries. Moreover, it is OS-portable and suitable for cloud software, such as Docker. We have preferred this database to experiment with Gremlin, due to both its good performance and its ease of installation and management.

**rdf2pg, architecture and approach**. Figures 1 and 2, show the mapping approach rdf2pg is based on. An RDF graph can be seen as a data model where part of the triples are about LPG node's plain properties and another part are about LPG relations, where one can map the latter from either straight RDF triples (i.e., which yield LPG relations with essentially any attribute) or from set of triples that correspond to the semantic of reified statements or similar entities. Such mappings, which are dataset or domain-specific, are defined by means of SPARQL queries. We selected SPARQL as the language for this mapping since it offers the significant advantage of not needing to learn anything new if one is already proficient in Semantic Web technologies. A mapping is defined by four types of SPARQL queries. Firstly, we have queries that select the URIs of those RDF nodes that are to be mapped to LPG nodes. Secondly, these URIs are used as parameters in RDF resource-centric queries that select/map node attributes (key/value properties and labels) from the RDF. Two other types of analogous queries are defined for the LPG relations: one query defines the relation identifying URI, its type and endpoints (as already mentioned, we support the common model where a relationship has one type only), and another URI-parameterised query allows for picking up the relation properties (again, a set of key/value pairs). The latter can be omitted when mapping plain RDF triples (in which case, the relation URI is a fictitious reference that can be built with appropriate SPARQL constructs, see [44] for an example).

In Figure 3 we show how the mapping is enacted by the rdf2pg architecture. For the case of nodes, a SPARQL query that selects node URIs is submitted to an input Jena TDB triple store, then, returned node URIs are batched and each batch is processed in parallel by an instance of a node handler. The node handler has a base abstract class, which has the common logic to do two things: fetching the node attributes from RDF, using the corresponding queries and, for each node, producing an abstract representation of the node's data (i.e., an abstract LPG view, see [45] for details). The abstract Node handler is then extended in the specific LPG data transformers, so that they can turn the abstract node representation into the representation that they need, e.g., Cypher CREATE instructions for the Neo4j converter or XML fragments for the GraphML converter. Conceptually similar components are available to produce LPG relations, that is, a relation processor that selects the relation URIs and types, and a relation handler, which produces LPG relations from a relation property selector defined in SPARQL. Clearly, a future extension to some other kind of similar RDF/LPG transformer would be based on this architecture and thus it would require its own implementation of the node handler and relation handler. As further mapping flexibility, the framework supports the definition of multiple query sets (sets of node and relation mapping queries), for those cases where a single RDF graph has subgraphs that are mapped differently to the LPG model. For instance, one might have to map gene and experiment representations, which are entities with enough differences (classes, node properties, relations) to warrant separated mappings.

## 3. Results

In this section, we evaluate rdf2pg as a tool to expose the conceptually equivalent data as multiple datasets, supporting multiple data access languages. This can be considered an expansion of a similar evaluation work initially done for the rdf2neo tool [13]. In particular, first, we will briefly review the three query languages that we have used and provide a qualitative analysis stemming from our experience and the lessons learned during the development of rdf2pg. Then, we present a quantitative evaluation of, on the one hand, the performance of rdf2pg in converting from RDF to three different graph databases, and on the other hand, the execution performance of semantically equivalent queries written in three different query languages, each against the rdf2pg-populated databases. Overall, this work aims at a broad comparison of essentially equivalent datasets managed with different graph databases and graph query languages, where their semantic alignment is obtained through a tool like rdf2pg.

### 3.1. Test datasets

Details about our benchmarks are available at the github repository [46]. As explained there, we have used three datasets about plant biology.

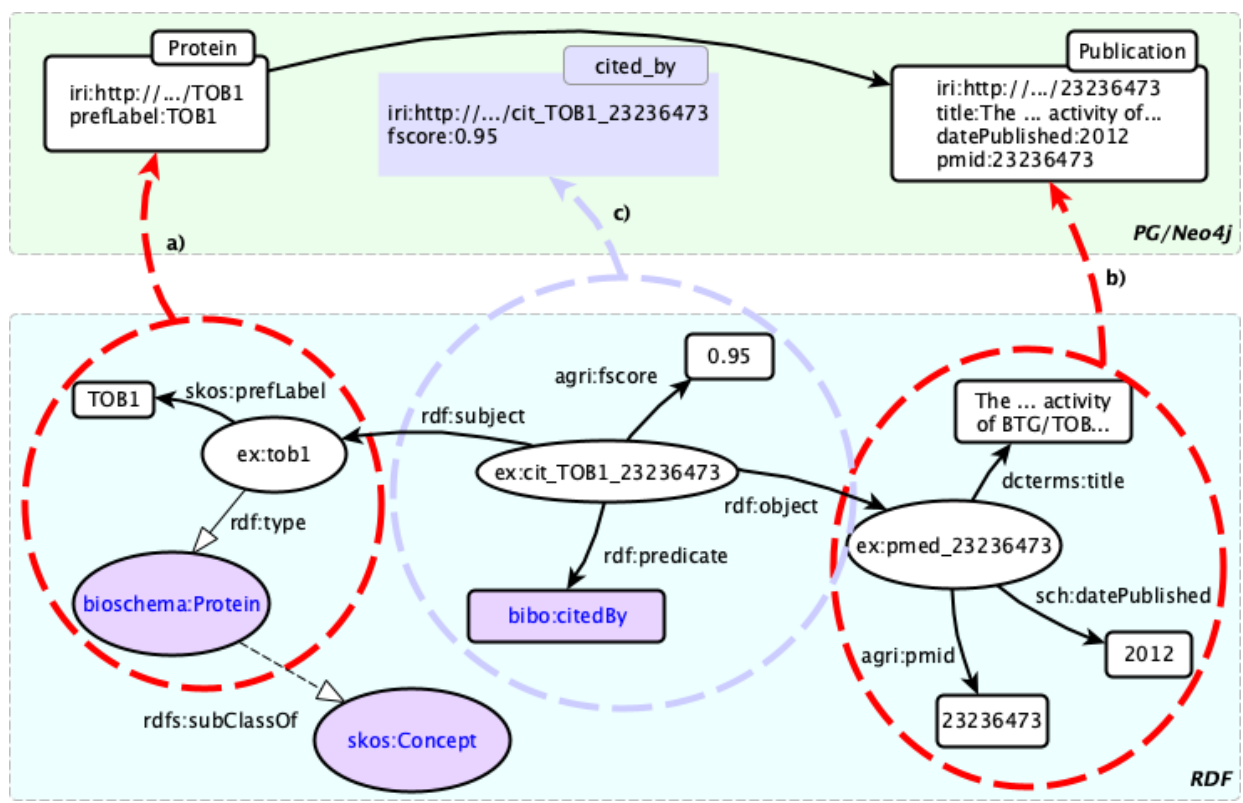

Fig. 1. An example (from [49]) of a Labelled Property Graph (top) and how it can be represented as RDF triples (bottom).

All the datasets have similar entities and mostly follow the same schema, which allows for assessing the scalability of the same queries. Figure 4 shows elements of such schema that we considered to write the benchmark queries. Essentially, one aspect these elements describe is gene annotations, including Gene Ontology [47] annotations, gene mutations (SNPs) and associated mutation phenotypes, together with phenotype annotations from the Plant Trait Ontology [48]. Another aspect is biochemical processes (so-called biological pathways) in which gene products are involved. The starting point for the benchmark was RDF files that represent these data, from which we populated Virtuoso triple stores (providing SPARQL access) by direct RDF loading, Neo4j databases (providing Cypher) using rdf2neo, and ArcadeDB databases (providing Gremlin) using rdf2graphml and then the Gremlin helpers to import from GraphML. The result for each dataset is that the three different databases contain data that are aligned to the same conceptual model, therefore it was possible to design several query tasks, which could be translated into semantically equivalent queries in the tested query languages (see our benchmark report for details about the semantic equivalence).

### 3.2. Qualitative considerations

In this section, we compare the three tested graph query languages, SPARQL, Cypher and Gremlin from qualitative perspectives. This includes examining the syntax and patterns that the languages offer, identifying which queries are easy to write and which ones present challenges. We present our experience in this area by referring to the queries used for the benchmark; all mentioned queries are listed in the benchmark code repository [46]. Both SPARQL and Cypher are declarative languages and both allow one to work with graph patterns, that is, templates that describe subgraphs in the database to be matched and retrieved. This declarative approach is typically exploited by query engines for optimisation operations, such as query rewriting.

```
a) b) Selection of node IRIs

SELECT ?iri {
  ?class rdfs:subClassOf* skos:Concept.
  ?iri a ?class.
}
```

```
a) b) Selection of node labels.
?iri is a named parameter

SELECT ?label {
  ?iri a ?label.
}
```

```
a) b) Selection of node properties (ie, attributes).
Works similarly for relations (e.g., see stato:score).
?iri is an instantiated parameter.

SELECT ?name ?value {
  { # Annotations of interest listed explicitly
    ?iri ?name ?value. VALUES ( ?name ) {
    (agri:pmid) (dcterms:title) ... }
  } UNION {
    # Annotations matching a criterion
    ?iri ?name ?value.
    ?name rdfs:subPropertyOf* schema:Property}
}
```

```
c) Selection of relation IRIs/types/endpoints,
reified relations

SELECT ?iri ?type ?fromIri ?toIri {
 ?iri a rdf:Statement;
    rdf:predicate ?type;
    rdf:subject ?fromIri;
    rdf:object ?toIri.
}
```

```
c) (cont.), selection of plain (property-less) triples

SELECT ?iri ?type ?fromIri ?toIri {
  # Relations of interest listed explicitly
  VALUES ( ?type ) {(dct:source)(sch:sourceOrganization)}
  ?fromIri ?type ?toIri.

  # each relation is assigned an ?iri in Cypher,
  # reified IRIs is a way to define them
  BIND(IRI(CONCAT(
    ex:, MD5(CONCAT( ?type, ?fromIri ?toIri ))))
    AS ?iri )
}
```

Fig. 2. How the RDF graph in Figure 1 can be mapped onto the corresponding LPG in rdf2pg tools, using SPARQL queries to list node or relation main elements, which are then passed to node-specific or relation-specific queries, to gather further element details

SPARQL, being bound to the RDF model, shapes this graph pattern paradigm around the idea of triple patterns (Figure 5), a network of nodes like the one in Figure 4 translates to a list of triples in the pattern, where nodes participating in multiple triples are simply listed with the same binding variable names. An advantage of this is that the syntax is rather simple and operations like querying integrated data (where entities with the same URIs were automatically merged in the database) are straightforward. Typical disadvantages are that certain patterns are rather verbose, eg, matching chains of nodes, matching many properties for a node, and dealing with data modelling workarounds such as reification might be even more verbose. The *joinRel* query from our benchmark is an example of the latter. On the other hand, Cypher is a language oriented to property graphs and its syntax is often more 'visual', feeling like you can draw the subgraphs to be matched, in particular chain patterns. For a typical example, compare the Cypher version of the *joinRel* example to the SPARQL one. An exception to this is in cases like *2UnionNest*, where graph patterns have to be built that involve many branches from hub nodes.

With past versions of Neo4j and Cypher, this was particularly hard to write, requiring to 'flow' partial results from one subquery to another (using the WITH and UNFOLD clauses), which is very unusual with respect to the more common UNION construct. Indeed, Cypher improved this construct recently (in version 4.0) and now such queries are easier to write, as we show in *2union1Nest+*.

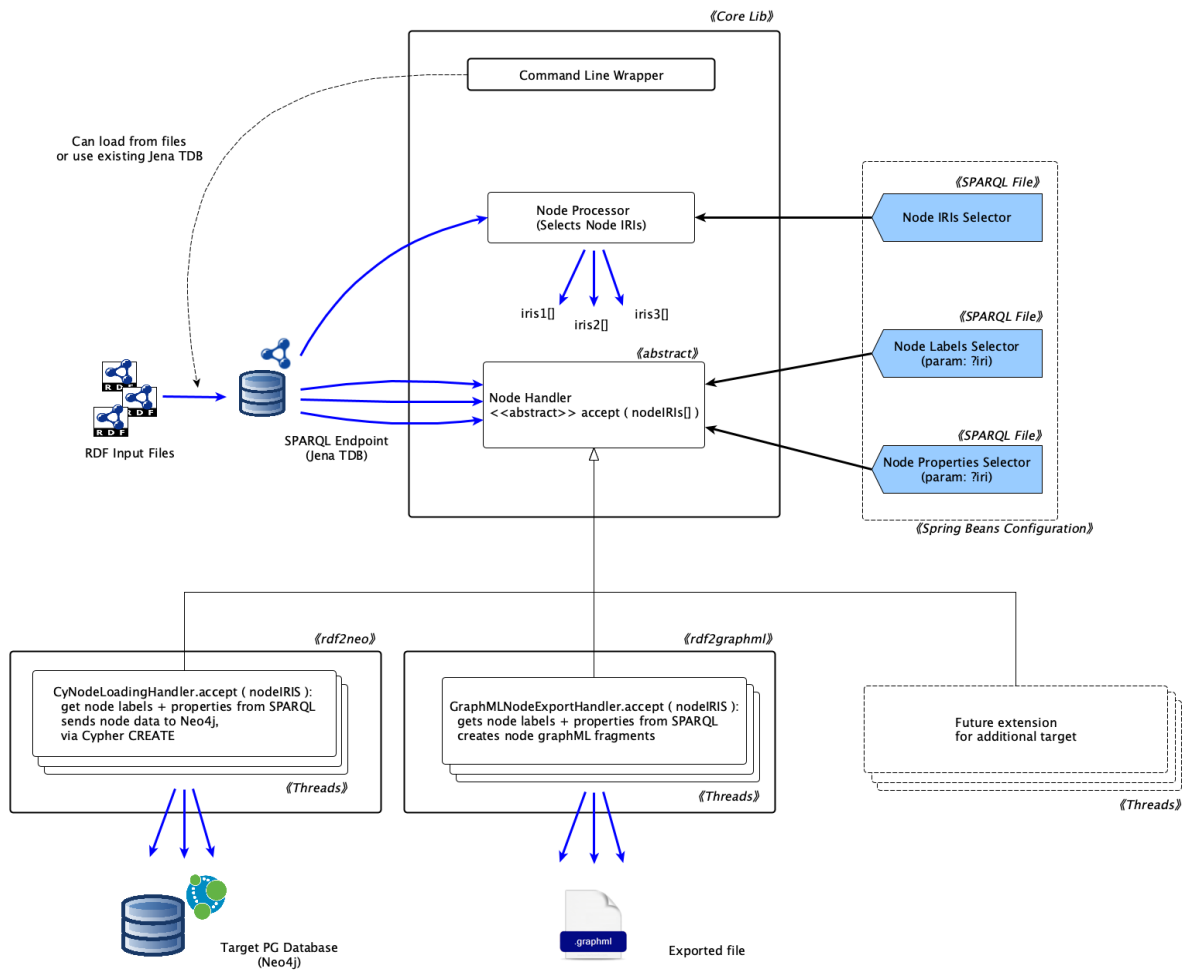

Fig. 3. : the rdf2pg architecture (from [50]). The diagram shows the components used to collect RDF batches of node references via SPARQL; the batches are then passed to parallel node handlers, which retrieve node details like properties (using additional SPARQL) and convert in-memory representations of LPG elements into a specific LPG target. Similar components are available to process relationships.

The Gremlin language contrasts with SPARQL and Cypher for giving the impression of a procedural language. In fact, writing a query looks similar to specifying the states/nodes and transitions/relations of a state machine. Common queries such as simple selections (see the 'selection' group in our benchmark) are as easy to write in Gremlin as in the other languages, and Gremlin is fairly simple when dealing with traversal patterns. Furthermore, the language is integrated with other programming languages like Groovy, which makes it a Turing-complete programming language [39], useful for advanced graph exploring tasks. For instance, our *joinReif* query defines a function to match a *uri* property in a relationship to a node (property graphs have data properties, but don't allow for links to other nodes or relationships). At the same time, this power often backfires. For instance, hub-based patterns (where many subgraphs depart from or join a node) are often hard to define in Gremlin efficiently, especially when the traversal paradigm makes it hard to link separated subgraphs through joining properties. Indeed, we needed to write the *joinReif* query using a lambda function, in order to make the *uri*-based join efficient.

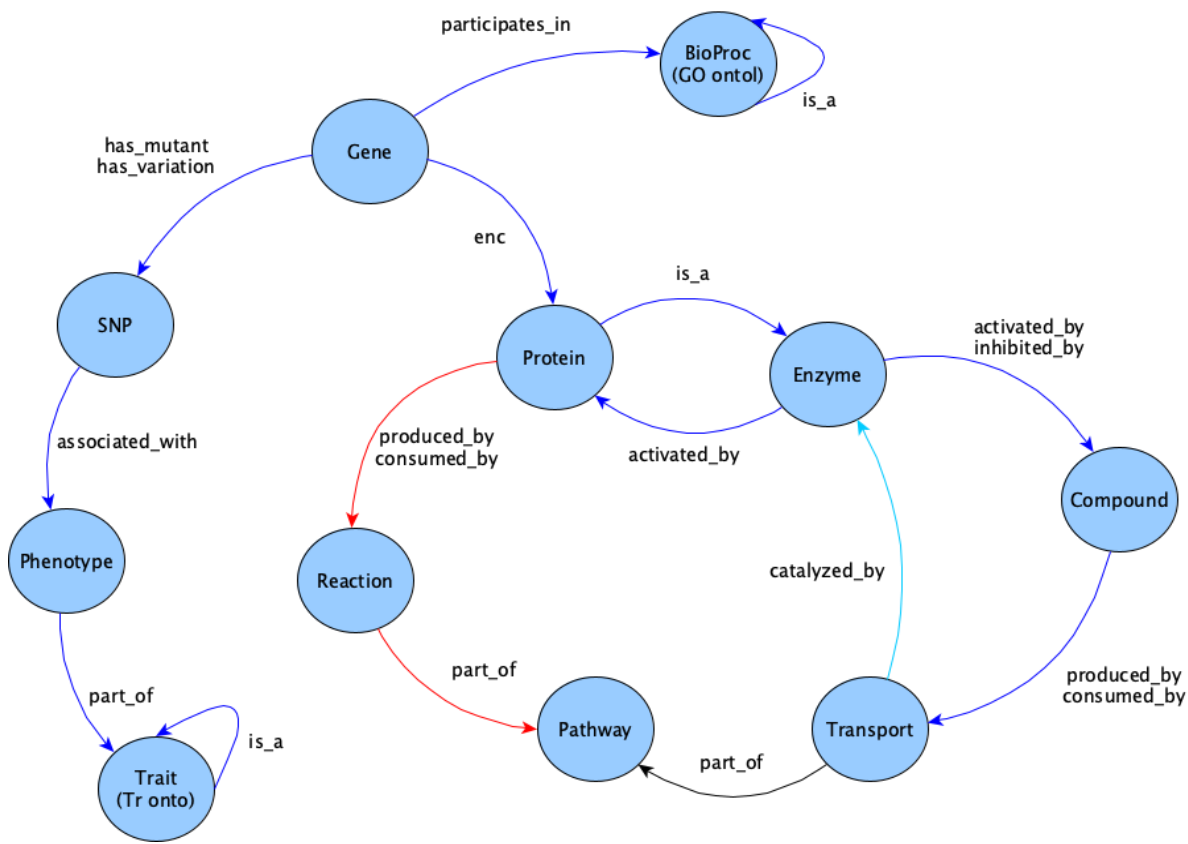

Fig. 4. typical elements of the plant biology datasets that we have used for benchmarking three different graph databases and query languages, with data coming from the same RDF sources and kept aligned using rdf2pg tools. Details in our dedicated code repository [46].

Related to this nature of mostly imperative query language, Gremlin leaves the query writer in control of how the graph database is explored, which can be very flexible, but at the same time, can force the query author to pay much more attention to query optimisation, including cases where such optimisation would be automatically computed by more declarative languages. We have experienced this in queries like *2union*, where traversing the graph in the order enzyme-to-protein is faster than protein-to-enzyme, due to the cardinality difference for the two types. For similar reasons, we have experienced significant challenges in writing aggregations using Gremlin. For example, *existAg* pinpoints biological pathways with certain characteristics and then computes both a reaction-per-pathway count and an average of proteins per reaction. This is a long query to write in SPARQL and Cypher, which becomes even longer and more convoluted in Gremlin, requiring a first part of traversals, which is combined with aggregating traversals starting from the elements projected from the first part. Another comparison we made concerns queries that traverse graph chains, ie, graph patterns like: x-to-y-to-z… (see the 'paths' category in our benchmark). The Cypher syntax is very compact and easy with this kind queries, and it is manageable to write them in SPARQL and Gremlin, although with more verbose patterns. These queries become more complicated when a traversal pattern has variable-length relations (eg, a pattern that matches both protein1/xref/protein2 and prot1/xref/prot2/xref/prot3). Both SPARQL and Cypher support this in their syntax, while Gremlin requires care with writing *repeat()* traversing steps (see our 'paths' queries and section 3.27 in [40]). Things become even more complicated for traversal paths with 'optional tails', eg, see the *lngSmf* query in the benchmark, where the chain tail consisting of protein/protein/gene is made optional. The Cypher syntax for defining such a case is still straightforward, since a chain tail having optional relations only (with min cardinality set to 0) is implicitly considered optional in its entirety, that is, the pattern doesn't need to match the nodes linked by such relations. Both

SPARQL and Gremlin require an OPTIONAL clause for achieving the same, since node matching in these languages is not optional in this case.

*3.3. Performance benchmark*

In this section, we show the results from a quantitative benchmark that we have run over the three chosen databases and languages, measuring the times taken to populate the graph databases, as well as the average times taken by the test queries to complete in multiple executions. For these tests, we have used: Virtuoso, Open Source edition, version 7.2.10-r16, Neo4j, Community edition version 5.11.0, ArcadeDB: version 23.4.1. We have used the rdf2pg framework and tools version 5.0. We have run these systems on a virtual machine equipped with: CPU, Intel(R) Xeon(R) Gold 6152, 2.10GHz, 8 cores, 32Gb RAM assigned to each DB server, 32GB assigned to rdf2pg tools. We have employed these settings to test three different datasets (referring to similar data and sharing a very similar schema) of increasing sizes of about 2, 21 and 97 million of RDF triples. Results and more details about the test settings and approach are described in the already-mentioned github repository [46].

**Database population performance**. The times taken to load the three test datasets into the three test databases are shown in Figure 6. This shows that even the largest dataset can be uploaded to any of the databases in a reasonable time (with a range between about 14 seconds to 30 minutes), at least when considering the case where a large dataset does not need frequent updates (eg, it is uploaded once a week or less often). We have also experienced that all the databases scale the uploading time linearly with the data size. In addition to the loading times, we have verified expected behaviours that depend on the details of rdf2pg. For example, the Neo4j population is influenced by the fact that the database is written by the rdf2neo tool, which has the overhead to read the input via SPARQL. In contrast, Virtuoso is the fastest database in the loading task, since it just needs to read RDF data and has optimised support for that. Though not within the scope of this paper, these figures could be improved by several optimisations. For example, the ArcadeDB population (and in general, any Gremlin-based writing) could be realised in a single step, where data are read from SPARQL and streamed to the target Gremlin database, similarly to the way the rdf2neo tool works. Adopting the performance-optimised RDF format HDT might be another improvement [51]. All of these hereby results are in line with our previous work [13].

**Query performance**. As explained in the benchmark report [46], we have designed 5 graph query categories and a total of 25 queries, based on both real use cases and other benchmark works, such as the Berlin benchmark [52]. As explained above, for each query, we wrote semantically aligned versions in the three tested languages. Results are shown in Figure 7. As shown, all the databases/languages perform within the order of hundreds of ms for most queries (each finding and fetching the first 100 records on average). In particular, Neo4j is the fastest database in many cases. Results also show that Virtuoso is still a good choice for pure-RDF and pure-SPARQL applications, although this triple store has the limitation of not supporting the above-mentioned RDF-star. Gremlin on top of ArcadeDB is often the slowest endpoint, this might depend on the fact that ArcadeDB is a relatively new product and its developers are still actively improving it. Moreover, Gremlin is usually implemented starting from a common code base and that might affect its performance compared to languages which are more native to a given database and its query engine. Another factor to consider is that, as said above, Gremlin is more sensitive to how queries are written. For instance, for the *join* and *joinRel* queries, we have noticed that the traversal step where returned results are limited to the first 100 matches matters for performance significantly, since exploring only the first 100 short chains in a graph pattern avoids that the engine traverses many more subgraphs and then cut the results to be returned at the end (it also might change the query semantics and results, see our github report for details. This has also an impact on the scalability towards the database size, in fact, while both Virtuoso and Neo4j show good scalability, this is more problematic with ArcadeDB. Considering specific queries, as expected, the fastest, most homogeneous and most scalable queries were selections/projections from simple patterns, while the aggregations were among the most challenging queries. This is in line with existing literature [13,52,53]: basic matching and projection are among the most used features in most query languages, while aggregations are notoriously hard to compute.

```
a) SPARQL

PREFIX rdf: <http://www.w3.org/1999/02/22-rdf-syntax-ns#>
PREFIX bk: <http://knetminer.org/data/rdf/terms/biokno/>

SELECT ?pname ?cname ?evidence
WHERE {
  ?p a bk:Protein;
     bk:prefName ?pname.

  ?rel a rdf:Statement;
       rdf:subject ?p;
       rdf:object ?cpx
       rdf:predicate bk:is_part_of.

  ?cpx a bk:Protcmplx;
       bk:prefName ?cname.

  ?rel a rdf:Statement;
       rdf:subject ?p;
       rdf:object ?cpx
       rdf:predicate bk:is_part_of.

  ?rel bk:evidence ?evidence.
}
```

```
b) Cypher

MATCH (p:Protein) - [rel:is_part_of] -> (cpx:Protcmplx)
RETURN p.prefName, cpx.prefName, rel.evidence
```

```
c) Gremlin

g.V().hasLabel ( 'Protein' ).as ('p')
 .outE().hasLabel ( 'is_part_of' ).as ( 'rel' )
 .outV ().hasLabel ( 'Protcmplx' ).as ( 'cpx' )
 .select ( 'p', 'cpx', 'rel' )
 .by ( 'prefName' )
 .by ( 'prefName' )
 .by ( 'evidence' )
```

Fig. 5. The graph query languages we tested, the example is a simplified version of the joinRel query in our benchmark [46] and based on the graph in Figure 4. a) SPARQL, syntax like bk:Protein shows the use of URIs to identify entities (resources in RDF jargon). The rdf:Statement match is an example of reification used to link the bk:evidence property to a part-of relationship (based on the standard rdf: vocabulary). b) Semantically equivalent query in Cypher, where a more visual and more compact syntax can be noticed. c) The Gremlin version, showing queries in this language describe graph traversals, by means of steps from one node or edge to another (V(), outV(), outE()) filtering steps (hasLabel()) and a number of other steps, such as selectors (select()), step modifiers (by()) and reference-creating steps (as()).

The semantic motif paths queries (i.e., a kind of chain pattern queries) are a particularly interesting category for the KnetMiner use cases, since in KnetMiner we often follow path chains to find entities associated with genes. As expected, Neo4j and Cypher excel in this kind of query, which is in line with their authors claiming their database engine is optimised for traversals. We were surprised these queries can be challenging for Gremlin, after further investigation, we noticed graph patterns with variable length relations (eg, find protein pairs linked by a chain of 'xref' relations of max length = 3) can become quite slow with large datasets. Again, this is likely to depend on the way the query is written and the fact Gremlin traversals are hard to optimise automatically. Finally, the queries in the 'counts' category have very varying performance across different databases and we presume this depends on the fact systems like Neo4j store summary data like the total number of nodes or relations, while other engines run the corresponding queries every time that such summaries are asked.

### 3.4. Theoretical considerations

It is useful to add theoretical analysis of the algorithms used by rdf2pg. In previous work (supplemental material in [13]), we have shown how the SPARQL-based mapping from RDF graphs to LPG graphs can be formalised by means of abstract algebra and we proved that the Cypher queries we generate in rdf2neo correspond to the mapped LPG. Similar reasoning can be done for the GraphML conversion, that is, it is possible to formally define how an LPG maps to the XML elements of the GraphML format and use that to prove that the LPG we build from SPARQL queries is correctly converted into GraphML. Namely, the proof is analogous to the proof of Theorem 1 in our mentioned work, which combines definitions 4 (RDF/LPG mapping), 5 (transformation induced by a mapping) and the GraphML semantics.

Furthermore, we have shown that the computational complexity of rdf2neo is dominated by the SPARQL mapping queries and it is PSPACE in the worst case, with a significant class of queries that can be reduced to LOGSPACE. Since the conversion from the LPG entities extracted from these queries to GraphML is linear with respect to the selected LPG nodes and relationships, this computational complexity holds for the GraphML converter too.

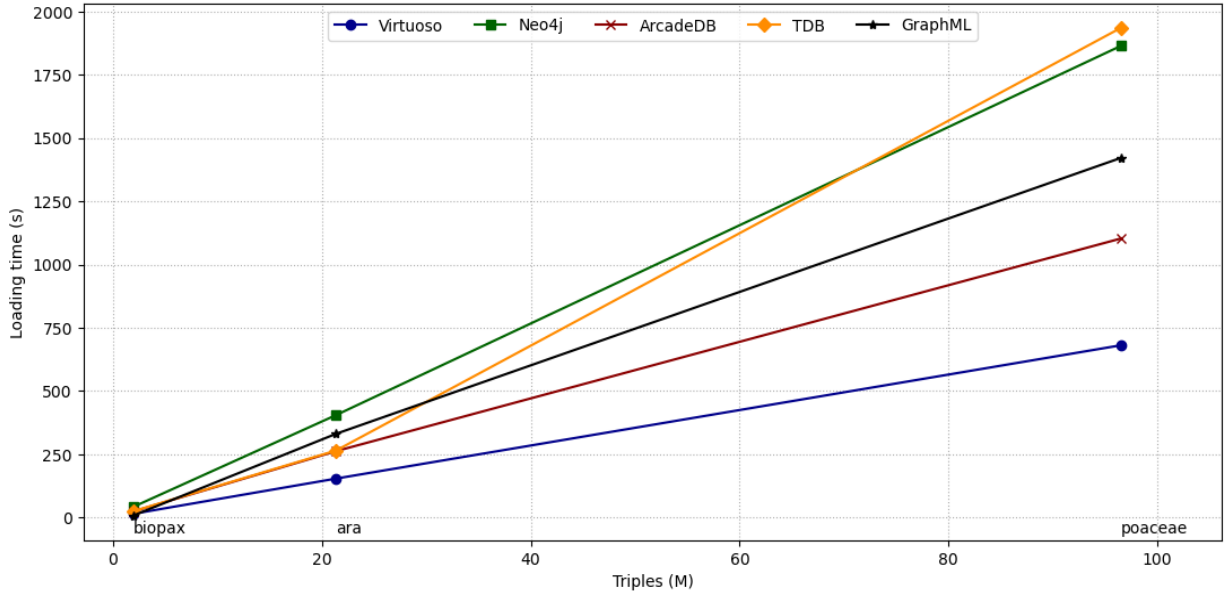

Fig. 6. *Results from graph database population with varying size datasets. More details at* [46].

In general, this computational complexity analysis is valid for any converter implemented with our framework, as long as the target-specific conversion has the same linearity (or does not exceed LOG-SPACE/PSPACE).

It is also worth considering more recent theoretical work concerning the properties of RDF/LPG mapping algorithms. Using the terminology introduced by [54], the SPARQL-based transformation we define by means of SPARQL (Definition 4 in [54]) is a kind of database mapping (from RDF to LPG), and such transformation uses SPARQL to also induce an LPG schema, consisting of all the labels, relation types, property domain and ranges that we implicitly extract from RDF. Such transformation/mapping is certainly computable (our tool computes it!) and it is semantics-preserving by construction, that is, the translated LPG is a valid instance of the schema constructed from SPARQL selections. According to the same mentioned paper by Angles et al., the information preservation of our transformation/mapping consists of the possibility of reconstructing the original RDF data from the translated property graph. In general, this is not a property of our RDF/PG transformations, eg, we might have instances of the classes 'Car' and 'Van' on the RDF side and, for some reason, one might want to define a mapping where all car and van nodes are assigned the 'Vehicle' label. Thus, in such a case, the original RDF data would be impossible to reconstruct (ie, no inverse PG/RDF mapping could exist). That said, we might show a set of conditions sufficient to make our transformation information-preserving. For instance, if the LPG node IRIs are always selected from a pattern like: *?iri rdf:type ex:Car*, and the label that is selected in this case is always *'Car'*, then, assuming no other selector maps to the *'Car'* label, all the *rdf:type* relations can be reconstructed correctly from the LPG labels. Similar conditions on relationships and node/relation properties could be defined to ensure this information-preserving property of our tool transformations. Note that, generally speaking, our SPARQL selectors consider a subset of the input RDF graph, hence this property of information preservation, that is, the possibility of going from the LPG data back to the RDF data the LPG was derived from, can only hold for such actually converted subset and not for the possibly bigger entire RDF graph one has started from. This corresponds to the real use of rdf2pg, where in most cases, the reversibility of RDF/LPG mapping is only interesting for the subset of data that are actually converted in either direction and moreover, such reversibility is not always a desired and sought-for property (eg, cars and vans merged into the vehicle class is a simplification where, likely, there is not interest in reconstructing the original RDF details).

## 4. Discussion

We have shown that labelled graph databases and their query languages have advantages and disadvantages that are complementary to more traditional Semantic Web technologies and Linked Data practices. The latter allows for managing dataset building and data sharing in a way that complies with the FAIR principles. In particular, RDF and existing ontologies or schemas based on RDF are still important means to ensure the goal of data interoperability. They also still play an important role in building pipelines to integrate heterogeneous data into unified knowledge graphs (i.e., ETL or ELT pipelines [55,56]). For instance, features like reusable URIs and standard schemas and ontologies produce graphs of data that are integrated in a seamless way.

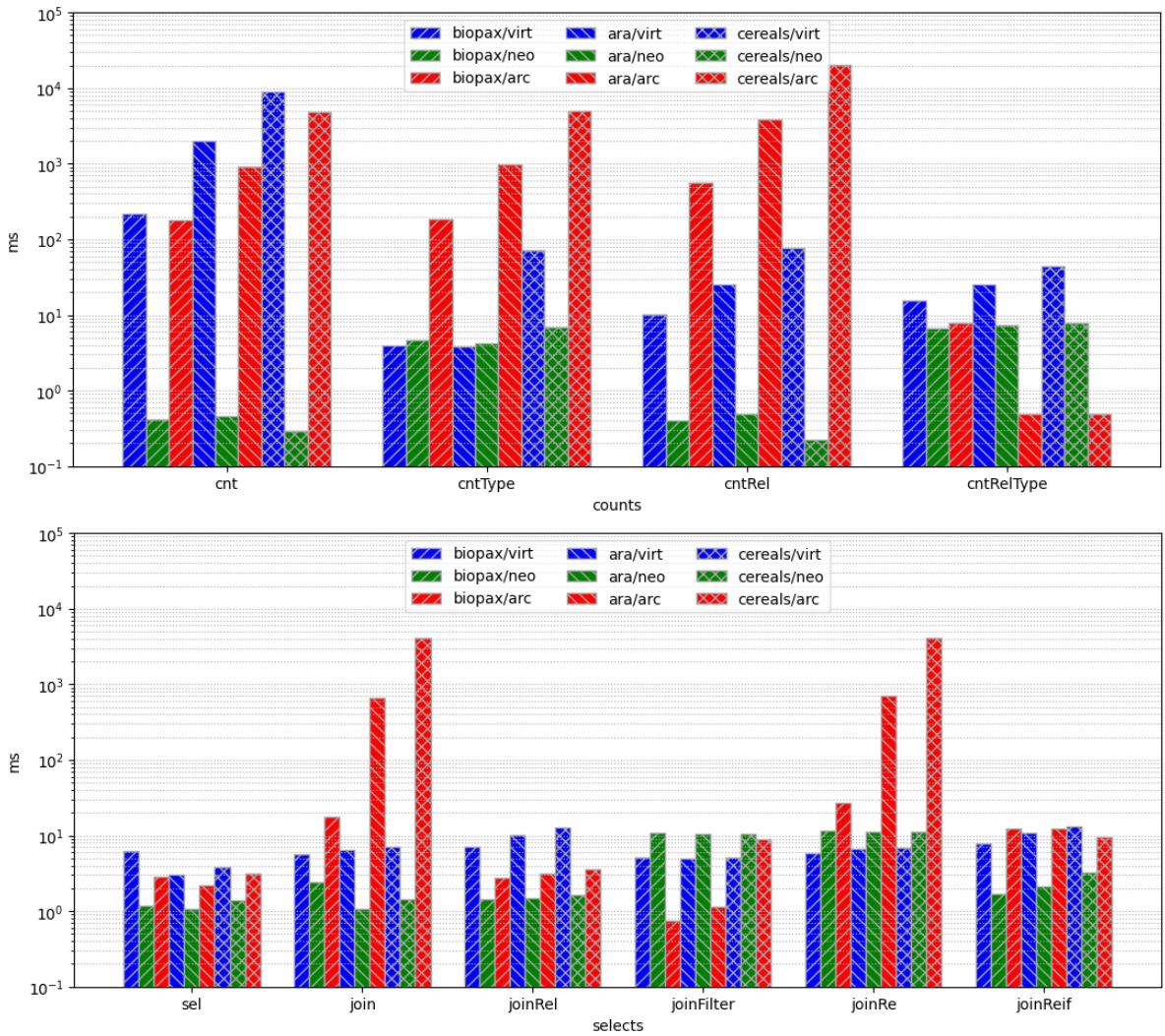

Fig. 7, Part 1. Results for benchmarking three various-size datasets on the three target graph query languages and graph databases, using queries of different categories. See [46] for a detailed list of all the tested queries, their description and code.

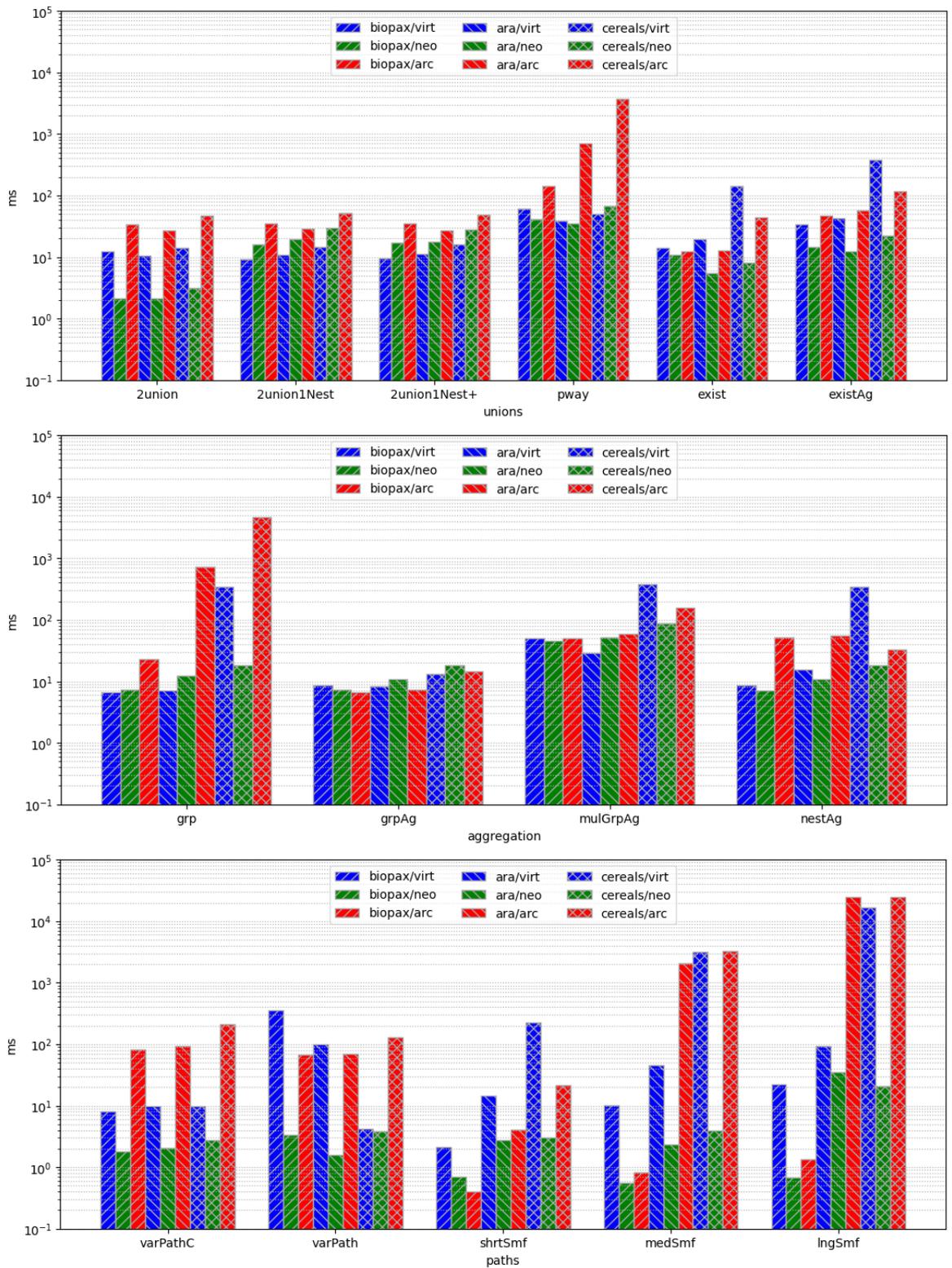

Fig. 7. Part 2.

On the other hand, users who are not proficient with SPARQL and the Semantic Web might prefer to query knowledge bases by means of languages like Cypher. We have shown that the performance of graph databases and triple stores are not extremely different, which allows for scenarios where data are first prepared using mainly RDF and related technologies and then loaded into a LPG database. Furthermore, our experience with KnetMiner proves that more complex architectures are feasible too, where, for instance, the same data are served via both SPARQL and Cypher, we could easily add Gremlin support and, at the same time, all the access points and encoding formats are aligned to the same conceptual data model. All of this is possible by means of the rdf2pg framework, which is both a base library to build RDF to property graph converters and two specific converters based on the framework. Our experience and our tests also show that, while query languages like SPARQL and Cypher have similar expressivity, Gremlin feels more like a lower level of abstraction and more suitable as a standard to build applications like multi-language or multi-model graph databases [42], or for use cases where advanced graph traversals are necessary.

### 4.1. Related work

As described in previous sections, this paper is an extension of our previous work on aligning RDF and Neo4j-based datasets, done within the context of the KnetMiner platform, where we have an interest in giving multiple access means to knowledge graphs [9]. Based on this initial work, we have seen the rdf2neo approach suitable for the generalisation and extensions described hereby and to be used to manage the enterprise ETL use case described above, where we have similar needs to align mixed data models and technologies. In developing our framework, we have relied on literature comparing the two types of graph paradigms. For instance, [57] discusses various definitions of knowledge graphs and their applications, [58] is a comprehensive review of knowledge graphs, how they are built and their applications, including link extraction from existing graphs. The above-mentioned work [54] gives formal definitions of graph databases and shows the kind of mappings that are possible between them. Work has been done to standardise property graph representations [59], convert between them automatically [14,60], or map query languages on multiple paradigms [61,62]. These conversion approaches are usually based on fixed mappings between the RDF and the LPG data model. For instance, all *rdf:type* relationships are converted into a node with a given label, all datatype triples become node properties and all node-to-node triples are turned into relationships. This makes the RDF-to-LPG conversion simple, since one does not have to design how to map their RDF model onto the corresponding LPG, which, for example, allows the Neosemantics tool for automatic back-conversion from Neo4j to RDF. However, a drawback of this pre-defined mapping is that it converts RDF graphs in a flat way, mostly ignoring the better expressivity of the LPG model. In particular, if a set of triples reifies a relationship with properties, they are mapped one-to-one on the LPG side, rather than producing the corresponding single relationship. In contrast, a major goal of our work is allowing for the definition of how RDF graph patterns should be translated onto LPG structures, especially in cases like reification. Although fixed mapping approaches avoid the overhead of defining custom mappings, our approach is more flexible and can address the cases where more natural (for the LPG model) mappings are desired. Moreover, we allow for using SPARQL as the language to define the mappings, since it does not require the users to learn any new special syntax, contrary to other approaches, such as [63]. Clearly, that is an advantage for Semantic Web experts, while it requires to learn at least the basics of RDF and SPARQL to other kind of data practitioners. Another limit of our approach is that it is not bi-directional, i.e., there is not an easy way to take a SPARQL-based mapping in the RDF-to-LPG direction and automatically compute the opposite LPG-to-RDF. This out of the scope of rdf2pg framework and possibly, it would need to be addressed with more declarative mapping languages (similarly to R2RML [64]). As mentioned, part of the queries we have used in the presented benchmark tests are inspired by the well-known Berlin benchmarks [52], a seminal work on RDF and SPARQL performance. Various other works on performance testing of relational and NoSQL databases exist, graph databases in particular. For example, [53] compared Neo4j and the relational database PostgreSQL, while [65] compared Cypher/Neo4, a Gremlin implementation on top of a Neo4j server, and a JPA object/relational mapping based on a MySQL database. In both cases, they found results similar to ours, that is, similar performance across the different storage and data access systems, with variability depending on the query types and use cases. Regarding the qualitative analysis of graph query languages, a recent study [66] surveyed different users of SPARQL and Cypher, concluding that they find them more similar than different. Interestingly, this contrasts with the analysis we have presented here, which highlights that user back-ground and expertise significantly influence their perception of Cypher, SPARQL, and their respective data models. Additionally, we identify expressivity differences in these languages that impact the ease of writing specific query types, such as multiple graph pattern hubs and long graph chain patterns, factors the cited authors did not consider in their analysis. The Gremlin and TinkerPop project started in 2009 as an Apache Foundation project, based on existing database graphs and models [39]. The idea of graph traversals stems mainly from work in the area of network analysis and graph processing algorithms [67]. To the best of our knowledge, our rdf2pg is the first that allows for customised mapping from RDF schemas to Gremlin-compatible databases and the first that compares the use of Gremlin and its performance to other LPG languages and their implementations.

## 5. Conclusions

Labelled property graphs are a strongly emerging approach to integrate heterogeneous data, with a number of graph databases and similar products supporting this data model. At the same time, Semantic Web technologies remain a reference to encode graph-like data in a standard way, which is open to any particular software system or data storage system. We have shown that different query languages to search graph data have complementary characteristics both in terms of usability and performance. Due to this, a framework like rdf2pg can help to efficiently convert RDF data onto labelled property graphs, choosing the LPG target that best suits one's needs and keeping data across different endpoints conceptually aligned. In fact, the major strength of rdf2pg consists in allowing for mapping from an RDF data schema to an LPG data schema via SPARQL patterns, which is often familiar to data management practitioners. The rdf2pg design offers the further advantage of making the framework easy to extend to new LPG targets. Indeed, we plan to support more of such extensions in future, for example, we are interested in supporting GQL as a future standard language to query graph databases [9]. SQL-like languages for graph databases are another category of LPG languages that we want to analyse in future, conducting a qualitative and quantitative evaluation similar to what we have presented in this paper. Related to this, we have done some preliminary work rewriting the benchmark queries with the Arcade DB SQL dialect, with first results showing the efficiency of this language [68]. Representing property graphs with an RDF schema is another ongoing standardisation effort [69,70] and we believe our tool would benefit from the adoption of such a standard, since it would allow for a further factorisation of the mapping that rdf2pg uses between an abstract LPG model and specific LPG targets. As stated previously, RDF-Star is a similar standardisation effort [32], which aims at extending RDF statements over statements. We look at this with interest as well, since it could be both a possible LPG target for rdf2pg or a richer RDF format to support as input.

## Acknowledgements

Funding: The contributions of MB, ADK and KHP to this work have been funded by the *Biotechnology and Biological Sciences Research Council* (BBSRC), through the projects of the Institute Strategic Programme: *Designing Future Wheat (BB/P016855/1)* and *Delivering Sustainable Wheats* (BB/X011003/1). The work of CB has been supported by Spanish national Project PID2020-113903RB-I00 (AEI / FEDER, UE), DGA / FEDER, and by the project *Semantic Data Management in Knowler*, Research Transference Project OTRI 2021/0432.

We would like to thank Kelvin Lawrence, for having published his excellent book on the Gremlin language and for his feedback related to the present work that he has given through the StackOverflow website.